\documentclass{iau}

\usepackage{amsmath}
\usepackage{graphicx}
\usepackage{multirow}

\newcommand{\rGc}{r_{\rm G}}

\begin{document}
\lefttitle{Di Cintio et al.}
\righttitle{Dynamical friction on Fornax}

\jnlPage{1}{4}
\jnlDoiYr{2025}
\doival{10.1017/xxxxx}

\aopheadtitle{Proceedings IAU Symposium}
\editors{Hyung Mok Lee, Rainer Spurzem and Jongsuk Hong, eds.}

\title{Mass loss and dynamical friction on the Fornax dSph galaxy in the Milky Way potential}

\author{Pierfrancesco Di Cintio$^{1,2,3}$, Giuliano Iorio$^{4,5}$, Carlo Nipoti$^6$, Francesco Calura$^7$}
\affiliation{$^1$CNR-Istituto dei Sistemi Complessi, Via Madonna del piano 10, I-50019 Sesto Fiorentino, Italy}
\affiliation{$^2$INFN-Firenze, via G.\ Sansone 1, I-50019 Sesto Fiorentino, Italy}
\affiliation{$^3$INAF-Arcetri, largo E.\ Fermi 5, I-50125, Firenze, Italy}
\affiliation{$^4$Institute of Cosmos Science, University of Barcellona, Martí i Franquès, 1 08028 Barcelona, Spain}
\affiliation{$^5$INAF-Padova, Vicolo dell’Osservatorio 5, I–35122 Padova, Italy}
\affiliation{$^6$Dipartimento di Fisica e Astronomia “Augusto Righi” – Alma Mater Studiorum – Università di Bologna, via Gobetti 93/2, I-40129 Bologna, Italy}
\affiliation{$^7$INAF-Bologna, via Gobetti 93/3, I-40129, Bologna, Italy}
\begin{abstract}
We study the interplay between mass-loss and dynamical friction (DF) on the orbital decay of the Fornax dwarf spheroidal galaxy in the potential of the Milky Way (MW). Using a simplified single particle approach combined with a mass-loss rate extrapolated by $N-$body simulations we find that the the effect of a time-dependent mass partially compensates DF, and typically produces a much less evident decay of the pergalactic distance, thus confirming that $N-$body simulations in smooth MW potentials without DF can be taken as a good model of the dynamics of dwarf satellite galaxies.  
\end{abstract}
\maketitle
\section{Introduction}
The Fornax dwarf spheroidal galaxy (dSph) has a remarkably high fraction of stellar mass locked in its globular clusters (GCs) system, thus putting at stake some GC formation models. It has been conjectured that an even larger fraction of stellar mass has been lost via tidal stripping.\\
\indent Aiming at constraining the amount of (stellar) mass lost by Fornax in \cite{2024A&A...690A..61D}, using high resolution $N-$body simulations combined with a a-posteriori mass component (stellar or dark matter) assignment scheme, we evolved the progenitor of the Fornax dSph along two Gaia DR3-based orbits in a static Milky Way (MW) potential. The core result of this work is that, even though, in principle, a broad range of initial stellar mass distributions for Fornax would be consistent with the present observed stellar density profile, the stellar mass loss must have been in all cases rather low (i.e. of the order of $\lesssim 3\%$). As a consequence, it appears rather unlikely that the mass loss only could explain the high fraction of stellar mass associated to Fornax's GCs system. Therefore the relative amount of field and GC stars remains a puzzle (despite the significant uncertainty in the adopted GC mass-to-light ratio) 
 for GC formation channels that assume originally much more massive progenitors.
\begin{figure}
  \includegraphics[width=0.85\textwidth]{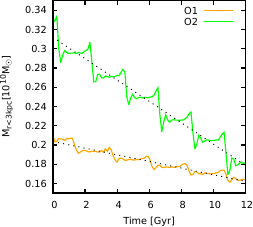}
\caption{Evolution of the mass within a control radius of 3 kpc for two different $N-$body simulations of the Fornax dSph in a smooth MW potential (coloured solid lines) and their linear fit (black dotted lines).}
\label{figca}       
\end{figure}
\section{Methods and models}
In line with \cite{2024A&A...690A..61D} here we assume the \cite{Joh95} MW model consisting of three components (disk, bulge and halo) with total gravitational potential $\Phi_{\rm MW,tot}(R,z)=\Phi_{\text{disk}}+\Phi_{\text{bulge}}+\Phi_{\text{halo}}$ composed by a \citet{MiyamotoNagai} disk
\begin{equation}
   \Phi_{\text{disk}}(R,z)= -\frac{GM_{\text{disk}}}{\sqrt{R^2 + \big(a+\sqrt{z^2 + b^2}\big)^{2}}}
   \label{eq:MijaNagai} 
\end{equation}
and a spherical \citet{Her90} bulge
\begin{equation}
    \Phi_{\text{bulge}}(\rGc)= -\frac{GM_{\text{bulge}}}{\rGc+c}
    \label{eq:JohnHern}
\end{equation}
embedded in a spherical \cite{1981MNRAS.196..455B} logarithmic halo
\begin{align}
     \Phi_{\text{halo}}(\rGc)= v^2_{\text{halo}} \ln{( \rGc^2 +d^2 )}.
     \label{eq:JohnHalo}
\end{align}
In the equations above $R=\sqrt{x^2+y^2}$ and $z$ are Galactocentric cylindrical coordinates, and $\rGc=\sqrt{R^2+z^2}$ is the Galactocentric distance, with $a=6.5$ kpc, $b=0.26$ kpc, $c=0.7$ kpc, $d=12$ kpc, $M_{\text{bulge}}=3.4\times 10^{10}\text{ M}_{\odot}$, $M_{\text{disk}}=10^{11}\text{ M}_{\odot}$ and $v_{\text{halo}}=128$ km/s.\\
\indent To include the effect of the \cite{1943ApJ....97..255C} DF on the orbit of the Fornax dSph, here we perform single particle integrations, where a test mass $M_{\rm tot}$ orbits in the MW potential according to the equations of motion
\begin{equation}\label{eomdynfric}
\ddot{\mathbf{r}}=-\nabla\Phi_{\rm MW,tot}(\mathbf{r})+\mathbf{a}_{DF},
\end{equation}
augmented by the deceleration produced by DF through 
\begin{equation}\label{dynfric}
\mathbf{a}_{\rm DF}=-2\pi G^2 M_{\rm tot}\rho(\mathbf{r})\ln(1+\Lambda^2)\Bigg[{\rm erf}(X)-\frac{2X^2}{\sqrt{\pi}}\exp(-X^2)\Bigg]\frac{\mathbf{v}}{v^3}.
\end{equation}
In the equation above $\rho(\mathbf{r})$ is the total MW mass density at $\mathbf{r}$, $\mathbf{v}=\dot{\mathbf{r}}$ and $X=v/(\sqrt{2}\sigma)$, where $\sigma(\mathbf{r})$ is the velocity dispersion at ${\mathbf{r}}$ (i.e. we assume a local Maxwellian approximation ignoring the MW velocity distribution function) as obtained by solving the Jeans equations for the MW potential and density (e.g. see \citealt{10.1111/j.1365-2966.2006.10404.x}). In Equation~(\ref{dynfric}), in order to take into account that the satellite is extended (e.g. see \citealt{2016MNRAS.463..858P}), we define the ratio $\Lambda$ entering the Coulomb logarithm as 
\begin{equation}
\Lambda={r_{\rm G}}/{{\rm max}\left [r_{\rm sat},\frac{G M_{\rm tot}}{v^2}\right]}, 
\end{equation}
where $r_{\rm G}$ and $r_{\rm sat}$ are the scale radii of the Galaxy and the satellite, which we fix to 12 and 4.3 kpc, respectively. $N-$body simulations performed in a static MW potential revealed that the mass loss rate experienced by the Fornax progenitor over 12 Gyr is nearly constant in time for a broad range of putative orbits. In Fig. \ref{figca} we show the evolution of the mass within a fixed radius of 3 kpc around the centre of Fornax for the orbits O1 and O2 discussed in \cite{2024A&A...690A..61D}. The dotted lines mark the linear best fits with the relation 
\begin{equation}
M(t)=M_0-\dot{M}t
\end{equation}
yielding the two mass loss rates $\dot{M}\approx3.4\times 10^{-3}$ and $1.1\times 10^{-2}[10^{10}M_{\odot}{\rm Gyr}^{-1}]$.
\section{Numerical simulations and discussion}
\begin{figure}
  \includegraphics[width=\textwidth]{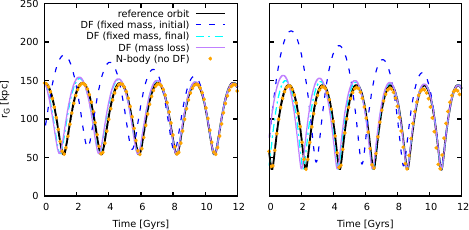}
\caption{Evolution of the Galactocentric distance $r_G$ of the Fornax dSph for two different models. The blue and cyan dashed lines refer to fixed mass single particle integrations with an artificially overestimated and a realistic present-day Fornax mass,  respectively $1.13\times 10^{10}{\rm M}_\odot$ and $5.7\times 10^{9}{\rm M}_\odot$. The purple solid lines refer to backward in time propagation where the mass loss has been accounted for. As a reference, the heavy black line marks the orbits propagated in the frozen MW potential without DF, and the yellow dots correspond to $N-$body simulations in the same frozen potential. Here $t=12$ Gyr is today.}
\label{xi}       
\end{figure}
We integrated backwards in time from $t=12$ Gyr to $t=0$ via Eq. (\ref{eomdynfric}) the orbit of a single particle representing Fornax using a modified leapfrog scheme (\citealt{mik06}) starting from the same two estimates of the present day phase-space coordinates used in our $N-$body simulations, with and without mass loss. In both cases we start assuming $M_{\rm tot}=5.7\times 10^9M_{\odot}$ at $t=12$ Gyr, that is the present-day total mass of Fornax as estimated in our $N-$body simulations. Additionally, we have also propagated from the same sets of coordinates a particle of constant mass $M_{\rm tot}=1.13\times 10^{10}{\rm M}_\odot$, corresponding to the heaviest Fornax progenitor used in our $N-$body study. Figure \ref{xi} shows the Galactocentric distance $r_{\rm G}$ as a function of time for the two choices of present day coordinates and the different simulation set-ups. It appears evident that allowing the mass to vary in time (purple lines, note that in backward integrations $M_{\rm tot}$ increases), in both simulations partially compensates the effect of DF, in particular with respect to the fixed mass runs with an overestimated present day mass (indicated by the blue dashed lines), thus implying a much lower apogalactic distance for the Fornax progenitor.\\
\indent For comparison, we have also reported the evolution of $r_{\rm G}$ in the $N-$body simulations (yellow dotted curves) where no MW DF was implemented. Remarkably, since the interplay between mass loss (nearly the 10\% for O1 and 40\% for O2) and DF results in a much milder orbital decay, we obtained further confirmation that our $N-$body simulations neglecting the DF exerted by the Galaxy can be safely taken as a reasonable model of Fornax evolution along a family of Gaia-based reference orbits. We note that, as shown in \cite{10.1093/mnras/staa1450}, under certain conditions the mass loss can sometimes enhance the orbital decay of a satellite system with respect to the simple point-particle with constant mass approximation. This is the case when extended tidal tails exert an additional torque on the satellite's orbit inducing the dissipation of orbital kinetic energy. This so-called tidal self-friction is therefore partially compensating the mitigation of the dtandard DF caused by mass loss. For the specific choices of initial conditions used in \cite{2024A&A...690A..61D} however, the tidal self-friction appears to be rather mild as no appreciable decay of the pergalactic distance was detected. In order to disentangle the effect of mass stripping and host galaxy DF one should in principle run $N-$body simulations in a galactic model incorporating DF, in a similar fashion to what has been carried out in \cite{2025arXiv250713904G}, where the sinking stellar system suffers the drag exerted by the host system on its center of mass. A numerical study of the strength of the DF force as a function of the satellite size and (possibly time dependent) mass is currently underway.  

\bibliographystyle{iaulike}
\bibliography{biblio_fnxgc} 
\end{document}